\documentclass[pla,reprint,onecolumn]{elsarticle}

\usepackage{hyperref}
\hypersetup{colorlinks=true, urlcolor=red, citecolor=blue, linkcolor=red}
\usepackage{graphicx}
\usepackage{subfigure}
\usepackage{epstopdf}
\usepackage{lastpage}

\begin{document}
\title{Nonclassicality generated by propagation of atoms through a cavity field}
\author{Arpita Chatterjee\corref{star}}
\ead{mailtoarpita@rediffmail.com}

\cortext[star]{Corresponding author, Tel.: +91-11-8527635535}
\date{\today}
\address{School of Physical Sciences, Jawaharlal Nehru
University, New Delhi 110067, India}

\begin{abstract}


We successively pass two $V$-type three-level atoms through a single-mode cavity field. Considering the field to be initially in a classical state, we evaluate various statistical properties such as the quasiprobability $Q$ function, Wigner distribution, Mandel's $Q$ parameter and normal squeezing of the resulted field. We notice that the sequential crossing of atoms induces nonclassicality into the character of a pure classical state (coherent field). The initial thermal field shows sub-Poissonian as well as squeezing property after interacting with the $V$ atoms.
\end{abstract}


\maketitle

\section{Introduction}

Generation and manipulation of nonclassical light field has been a major field of interest in quantum optics and quantum information processing \cite{bouwmeester00}. The study of these states provides a fundamental understanding of quantum fluctuations and opens a new way of quantum communication or imaging beating the standard quantum noise limit. Also nonclassical states have many real life applications.
For example, squeezed states are used to reduce the noise level in one of the phase-space quadratures below the quantum limit
\cite{scully97}, entangled states produced in down-conversion process are employed to test fundamental quantum features such as non-locality \cite{zeilinger98} and to realize quantum information transmission schemes (cryptography \cite{braunstein05,naik00} or teleportation \cite{bouwmeester98}). Quantum superpositions of fields with different classical parameters are used to explore the quantum or classical boundary and the decoherence phenomenon \cite{brune96}.

In this context, theoreticians as well as experimentalists have proposed various schemes to prepare nonclassical states of optical field.
Among them, subtracting photons from and/or adding photons to traditional quantum states provide an useful way to generate nonclassical state. Agarwal and Tara \cite{agarwal91} first proposed a method for producing the photon-added coherent state. Another way of creating photon-added or photon-subtracted state is through a beam-splitter \cite{ban96}. Dakna \cite{dakna98} showed that if the initial state and a Fock state are injected at the two input channels, then the photon number counting of the output Fock state reduces the other output channel into a corresponding photon-added or photon-subtracted state. The photon-added coherent states allow one to witness the gradual change from the spontaneous to the stimulated regimes
of light emission \cite{zavatta04}. Moreover, photon subtraction can be applied to improve entanglement between Gaussian states \cite{ourjoumtsev07,browne03}, loophole-free tests of Bell's inequality \cite{garcia04,nha04} and quantum computing \cite{bartlett02}.

Single-photon Fock state, a nonclassical state, is an indispensable resource in an all-optical quantum information processing device \cite{knill01}. These states can be prepared by controlling the emission of a single radiator: molecule \cite{brunel99} or quantum dot \cite{friberg92}. In addition, cavity QED experiments in which atoms interact one at a time with a high Q resonator can be used for Fock state preparation. A one-photon Fock state is created in this way by a $\pi$ quantum Rabi pulse in a microwave cavity \cite{maitre97} or by an adiabatic passage sequence in an optical cavity \cite{henrich00}. We report here how the passage of two $V$-type \cite{abdel07} three-level atoms transfers the classical cavity field into a nonclassical one.

We consider here a very basic model to describe the interaction of the quantum field with the atom after letting two $V$-type three-level atoms successively pass through it. This fundamental structure can be framed by adopting a Lindbladian point of view. We can correspond this model to a very simple situation, where a primary system interacts with a bath of harmonic oscillators at zero temperature, with an interaction Hamiltonian that resembles the Jaynes-Cummings format. We can start with the Born-Markov equation and tracing out the bath degrees of freedom, we can obtain an equation in the Lindblad form \cite{brasil11}. This interaction causes additional decoherence which can also be treated by using the Lindbladian approach \cite{nha05}.

This paper is structured as follows: we describe our problem and derive the wave function for the considered atom-cavity system in Sec.~\ref{sec2}. Sec.~\ref{sec3} concerns with finding the quasiprobability functions of the field left in the cavity. In Sec.~\ref{sec4}, we study the Mandel's $Q$ parameter and normal squeezing by taking the quantized initial field in a coherent or thermal state. The last section ends with a summary of the main results of this article.

\section{State Vector}
\label{sec2}

\begin{figure}[h]
\centering
\includegraphics[width=5cm]{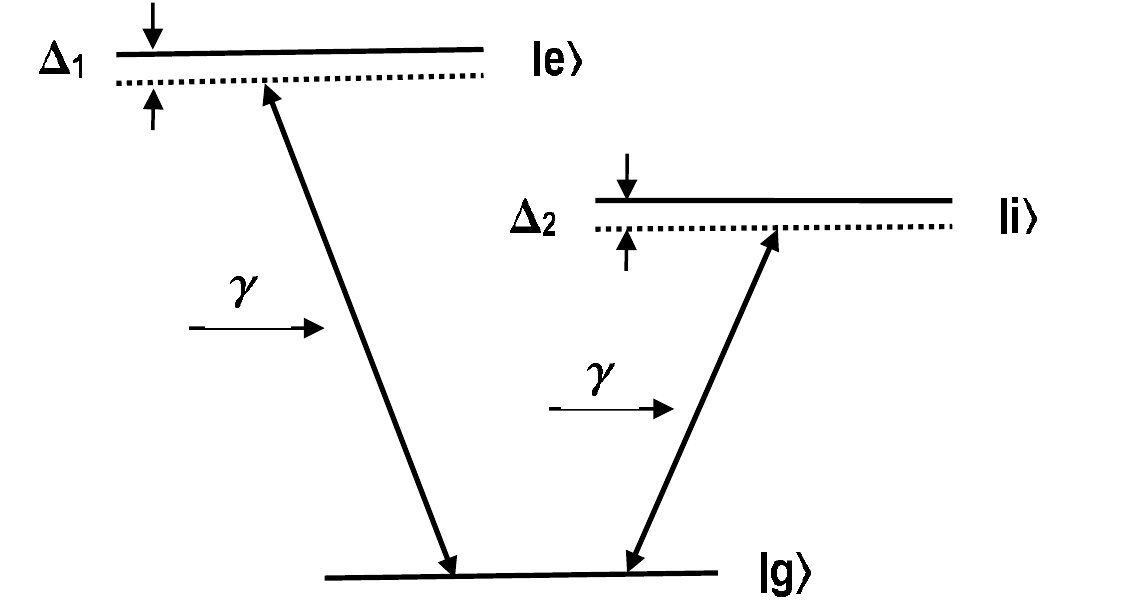}
\caption{Three-level atom in the vee-type configuration.}
\label{fig1}
\end{figure}

We begin by considering a $V$-type three-level atom having its higher-energy state $|e\rangle$ with energy $\omega_e$, intermediate-energy state $|i\rangle$ with energy $\omega_i$ and ground-energy state $|g\rangle$ with energy $\omega_g$. The atom interacts with a single-mode cavity field of frequency $\gamma$. $\Delta_1~(= \omega_e-\omega_g-\gamma)$ and $\Delta_2~(= \omega_i-\omega_g-\gamma)$ represent the respective detunings of the transitions $|e\rangle\leftrightarrow|g\rangle$ and $|i\rangle\leftrightarrow|g\rangle$ from the field mode as shown in Fig.~\ref{fig1}.
The \textit{RWA} leads to the interaction Hamiltonian (with $\hbar=1$) \cite{liu00}
\begin{eqnarray}\nonumber
H_{int} & = & g_1(ae^{i\Delta_1 t}\vert e\rangle\langle g\vert+a^\dag e^{-i\Delta_1 t}\vert g\rangle\langle e\vert)\\
& & +g_2(ae^{i\Delta_2 t}\vert i\rangle\langle g\vert+a^\dag e^{-i\Delta_2 t}\vert g\rangle\langle i\vert),
\label{eq1}
\end{eqnarray}
where $g_1$, $g_2$ are the atom-field coupling constants, $a\,(a^\dag)$ are annihilation (creation) operators for the field with canonical commutation relation $[a, a^\dag] = 1$, The solution of the Schr\"{o}dinger equation with Hamiltonian (\ref{eq1}) gives the state vector $|\psi(t_1)\rangle$ at any time $t_1\geq0$ for the coupled atom-cavity system. We assume that the atom enters the cavity in a coherent superposition of its eigenkets $\vert e\rangle$ and $\vert i\rangle$ that means $|\psi_a(0)\rangle=\frac{1}{\sqrt{2}}(|e\rangle+|i\rangle)$ and if initially the field is in the superposition of photon number states i.e. $|\psi_f(0)\rangle=\sum F_n|n\rangle$ with $\sum_n|F_n^2|=1$ then after evolution, the state vector of the considered atom-field system becomes \cite{peng92}
\begin{eqnarray}
|\psi(t_1)\rangle = \sum_n\left[c_{e,n-1}|e, n-1\rangle+c_{i,n-1}|i, n-1\rangle+c_{g,n}|g, n\rangle\right],
\end{eqnarray}
where
\begin{equation}
\left.
\begin{array}{lcl}
c_{e, n-1} & = & -g_1\sqrt{n}B_1\left[\frac{e^{i(\Delta/2+\beta_1)t_1}-1}{(\Delta/2+\beta_1)}-\frac{e^{i(\Delta/2-\beta_1)t_1}-1}{(\Delta/2-\beta_1)}\right]+\frac{1}{\sqrt{2}}F_{n-1},\\\\
c_{i, n-1} & = & -g_2\sqrt{n}B_1\left[\frac{e^{i(\Delta/2+\beta_1)t_1}-1}{(\Delta/2+\beta_1)}-\frac{e^{i(\Delta/2-\beta_1)t_1}-1}{(\Delta/2-\beta_1)}\right]+\frac{1}{\sqrt{2}}F_{n-1},\\\\
c_{g, n} & = & B_1\left[e^{-i(\Delta/2-\beta_1)t_1}-e^{-i(\Delta/2+\beta_1)t_1}\right],
\end{array}
\right\}
\end{equation}
with $\Delta_1=\Delta_2=\Delta$ and
\begin{equation}
\left.
\begin{array}{lcl}
B_1 & = & -\sqrt{\frac{n}{2}}\frac{(g_1+g_2)}{2\beta_1}F_n,\\\\
\beta_1 & = & \sqrt{\Delta^2/4+n(g_1^2+g_2^2)}
\end{array}
\right\}
\end{equation}
Later we assume that after interacting with the cavity field for time $t_1$ the atom exits the cavity in its ground state only \cite{pathak05,gerry96}. Then
\begin{eqnarray}
|\psi(t_1)\rangle = \sum c_{g, n}(t_1)|g, n\rangle.
\end{eqnarray}

Next we perform the transit of a second identical atom through the cavity. Like the previous one, this atom also enters the cavity in the state $\frac{1}{\sqrt{2}}(|e\rangle+|i\rangle)$ and stays there for time $t_2$.  Then for $g_1=g_2=g$ and for zero detuning, the system evolves to
\begin{eqnarray}\nonumber
\vert\psi(t_2)\rangle & = & \sum_n\left[D_{e, n-1}(t_2)|e, n-1\rangle+D_{i, n-1}(t_2)|i, n-1\rangle+D_{g, n}(t_2)|g, n\rangle\right],\\
\label{eq3}
\end{eqnarray}
where
\begin{equation}
\left.
\begin{array}{lcl}
D_{e, n-1}(t_2) & = & \frac{1}{\sqrt{2}}c_{g, n}(t_1)\cos(\sqrt{2n}gt_2),\\\\
D_{i, n-1}(t_2) & = & \frac{1}{\sqrt{2}}c_{g, n}(t_1)\cos(\sqrt{2n}gt_2),\\\\
D_{g, n}(t_2) & = &-F_n \sin(\sqrt{2n}gt_1) \sin(\sqrt{2n}gt_2).
\end{array}
\right\}
\end{equation}
This state vector $|\psi(t_2)\rangle$ describes the time evolution of the whole atom-field system but we now concentrate
on some statistical properties of the single-mode field. The field inside the cavity after departing the second atom is obtained
by tracing out the atomic part of $\rho(t_2) = \vert\psi(t_2)\rangle\langle\psi(t_2)\vert$ as
\begin{eqnarray}
\rho_f(t_2) = Tr_a[\rho(t_2)],
\label{eq8}
\end{eqnarray}
where we have used the subscript $a\,(f)$ to denote the atom (field).

This $\rho_f(t_2)$ will be of consideration throughout the next section to determine the statistical properties of the field left into the cavity.

\section{Quasiprobability function}
\label{sec3}

The quasiprobability distribution functions are important for the statistical description of a quantum mechanical
state in phase space. As the position and momentum cannot be defined simultaneously with infinite precision, the description of a
quantum mechanical state in phase space is not unique; there is a family of quasiprobabilities of which the Glauber-Sudarshan $P$,
Husimi $Q$ and Wigner functions are quite well-known. But most of the quasiprobabilities involve troublesome integrations over the phase
space variables. The exception is the $Q$ function, a coherent expectation of the field density matrix, and is therefore widely used to describe
the field dynamics in situations where the density matrix can be computed easily.

\subsection{$Q$ function}

The $Q$ function for the resulted cavity field is defined as the diagonal elements of the density matrix $\rho_{f}$ in the coherent state basis \cite{buzek93}
\begin{eqnarray}
Q(\alpha, \alpha^*) = \frac{1}{\pi}\langle \alpha|\rho_{f}(t_2)|\alpha\rangle.
\end{eqnarray}
Given a coherent state $|\alpha_0\rangle$ or a thermal field $\rho_{thm}=\sum_{n=0}^\infty\frac{{\bar{n}}^n}{(\bar{n}+1)^{n+1}}|n\rangle\langle n|$ as an initial state, the output state after interaction possesses the $Q$ functions respectively
\begin{eqnarray}\nonumber
Q_{coh}(\alpha) & = & \frac{1}{\pi}e^{-(|\alpha_0|^2+|\alpha|^2)}\left[\sum_{n, m}\frac{(\alpha_0\alpha^*)^{n-1}
(\alpha_0^*\alpha)^{m-1}}{(n-1)!(m-1)!}\sin(\sqrt{2n}gt_1)\cos(\sqrt{2n}gt_2)\right.\\\nonumber
& & \left.\sin(\sqrt{2m}gt_1)\cos(\sqrt{2m}gt_2)+\sum_{n, m} \frac{(\alpha_0\alpha^*)^n(\alpha_0^*\alpha)^m}{n!m!}\sin(\sqrt{2n}gt_1)
\right.\\
& & \left.\sin(\sqrt{2n}gt_2)\sin(\sqrt{2m}gt_1)\sin(\sqrt{2m}gt_2)\right],
\end{eqnarray}
and
\begin{eqnarray}\nonumber
Q_{thm}(\alpha) & = & \frac{1}{\pi}\frac{e^{-|\alpha|^2}}{(\bar{n}+1)}
\left[\sum_n\frac{1}{(n-1)!}\left(\frac{\bar{n}|\alpha|^2}
{\bar{n}+1}\right)^{n-1} \sin^2(\sqrt{2n}gt_1)\cos^2(\sqrt{2n}gt_2)\right.\\
& & +\left.\sum_n\frac{1}{n!}\left(\frac{\bar{n}|\alpha|^2}{\bar{n}+1}\right)^n \sin^2(\sqrt{2n}gt_1)\sin^2(\sqrt{2n}gt_2)\right].
\label{eq11}
\end{eqnarray}
\begin{figure*}[ht]
\centering
\includegraphics[width=7cm]{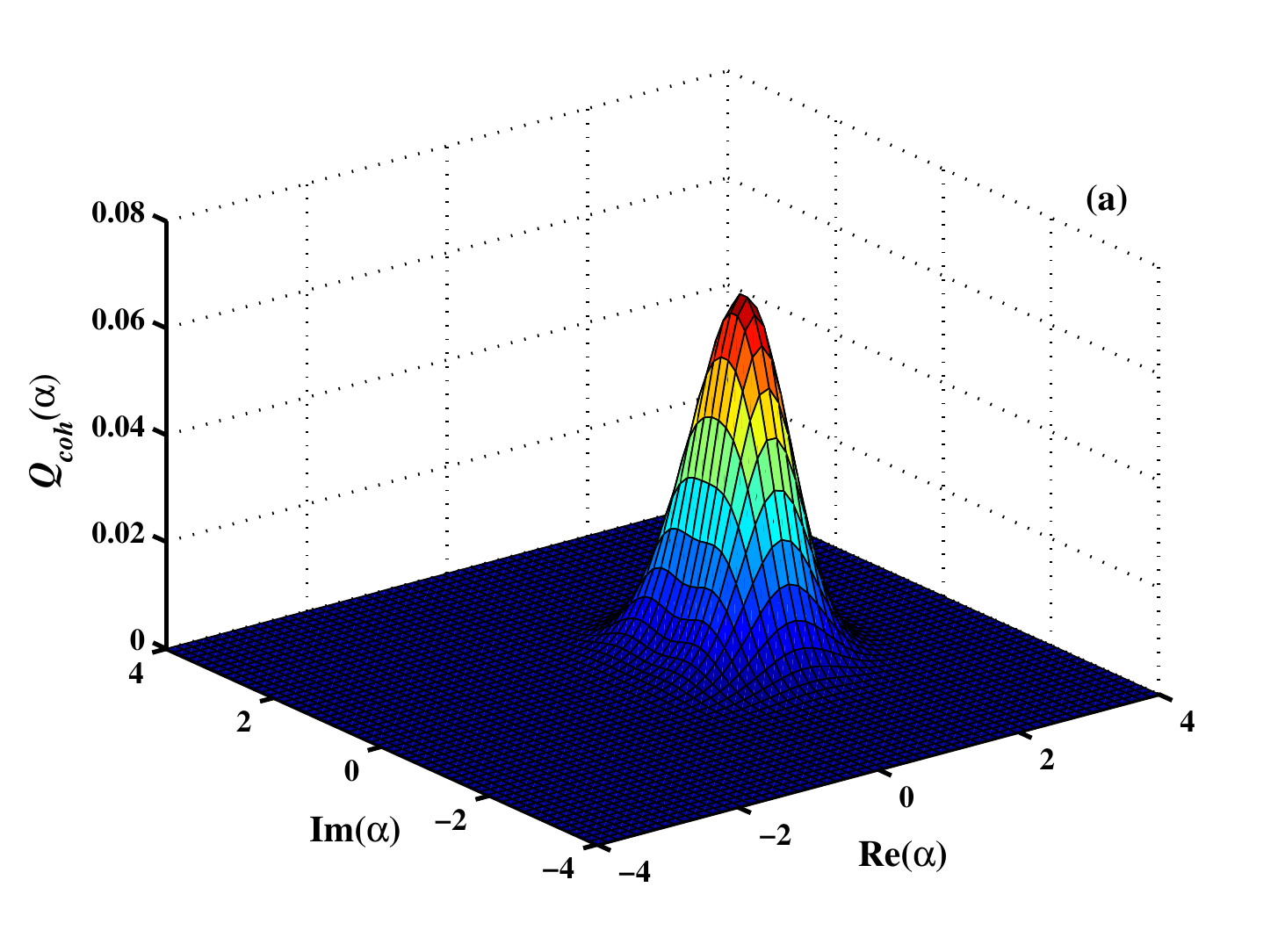}\hspace{2cm}
\includegraphics[width=7cm]{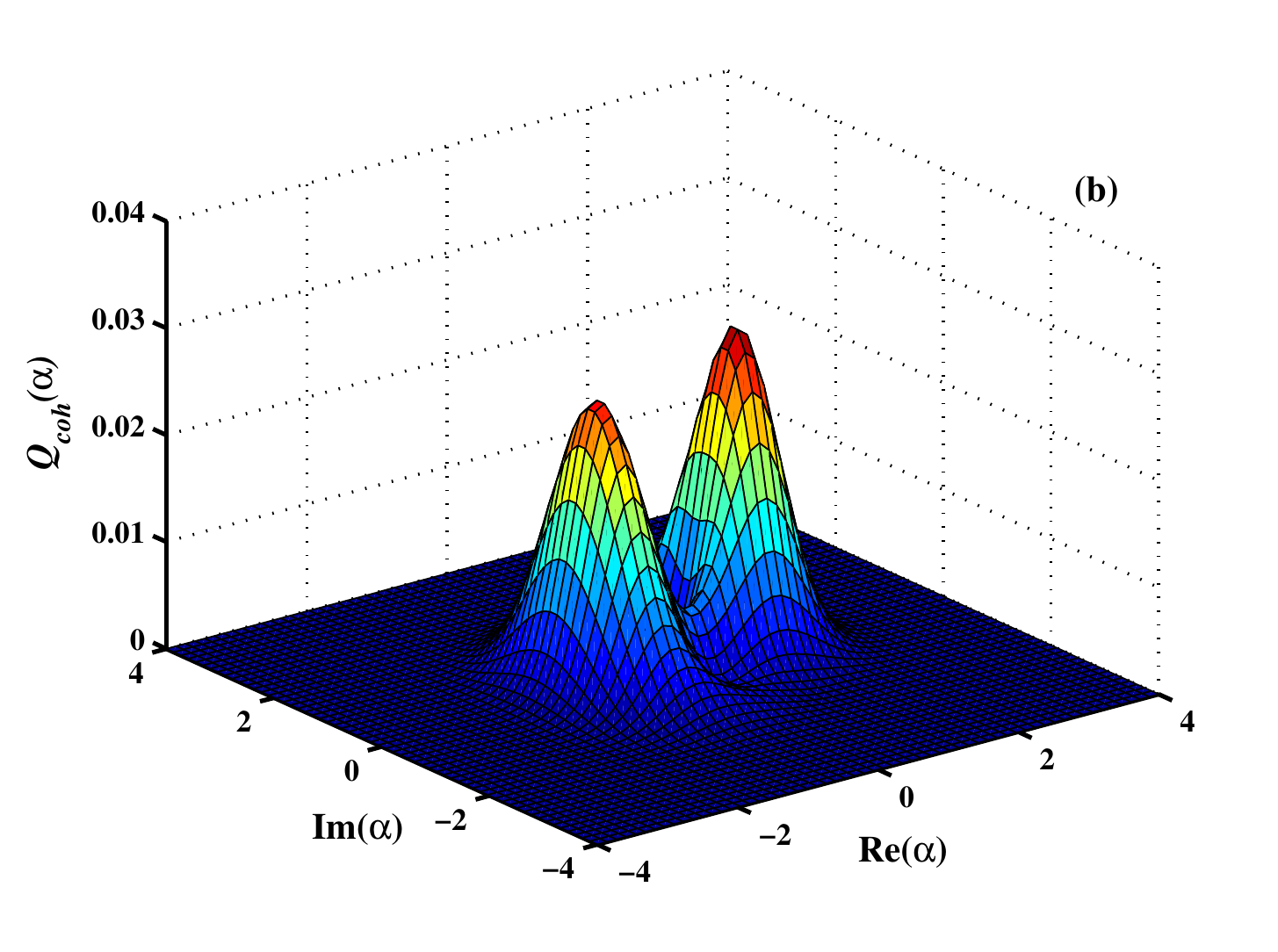}\hspace{2cm}
\includegraphics[width=7cm]{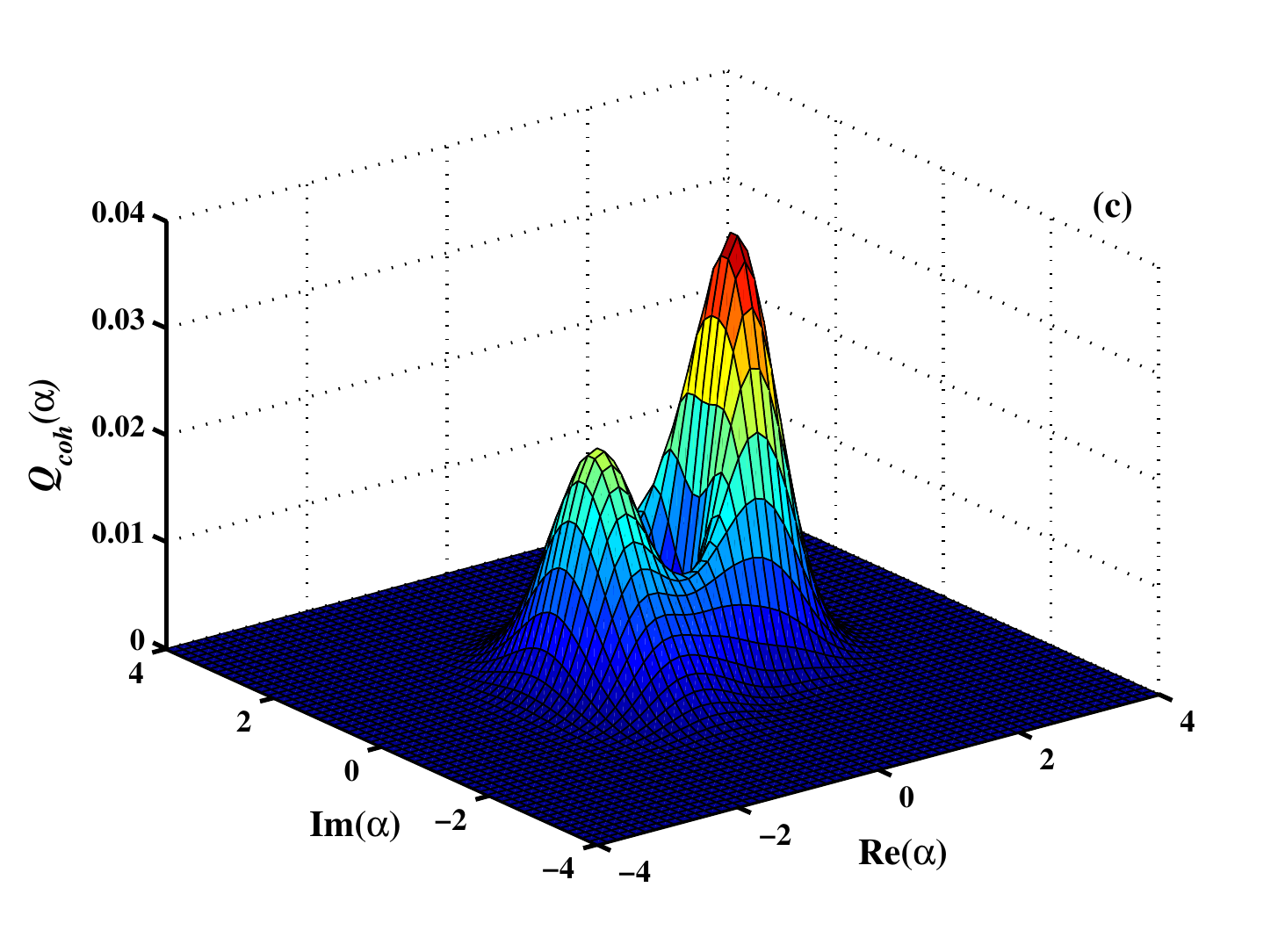}\hspace{2cm}
\includegraphics[width=7cm]{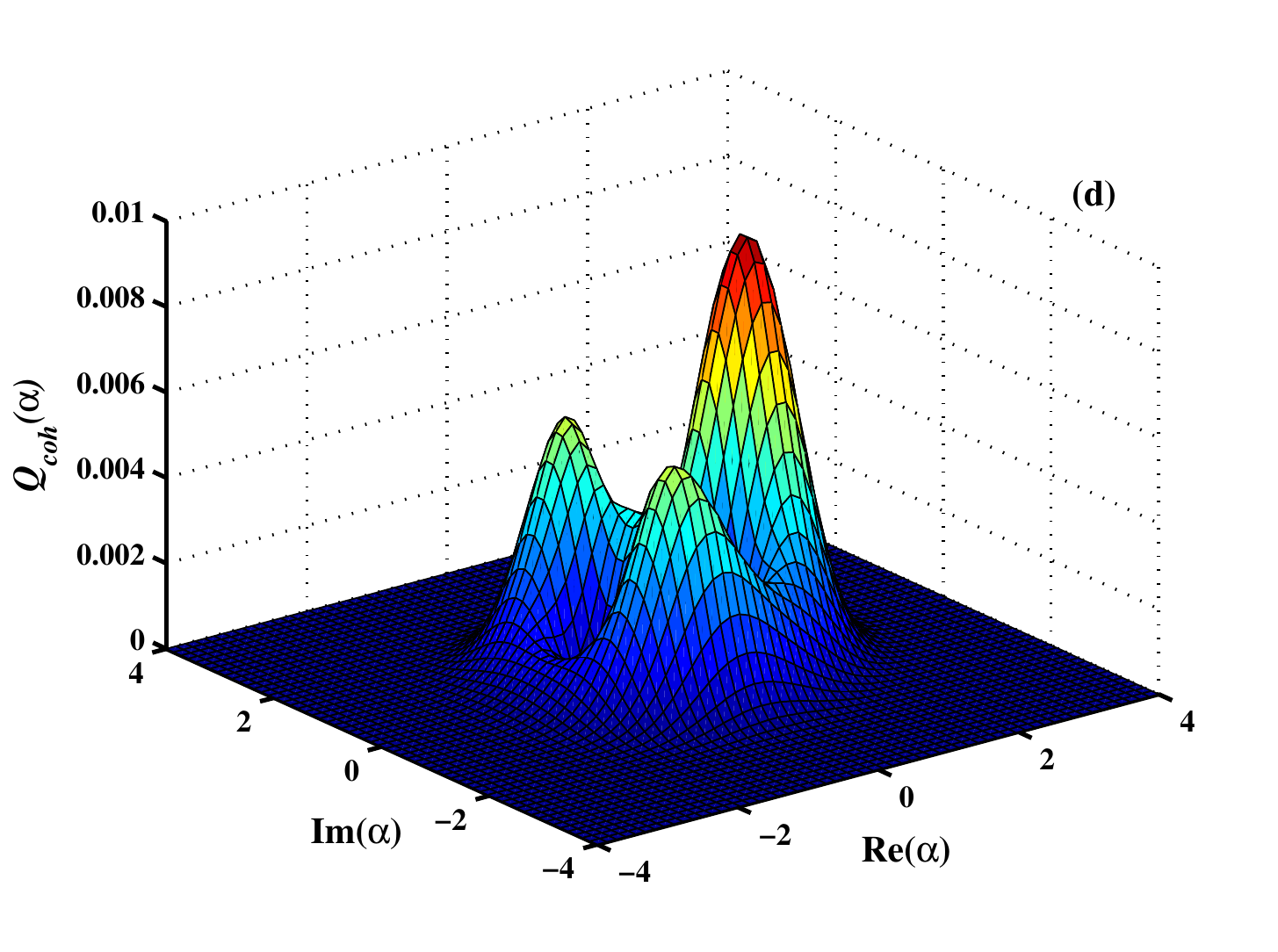}

\caption{(Color online) A graph showing the $Q$ function of the final cavity field for the initial coherent state with $\alpha_0=2$
at different times (a) $gt_1=gt_2=\pi/6$, (b) $gt_1=gt_2=\pi$, (c) $gt_1=gt_2=7\pi/6$ and (d) $gt_1=gt_2=2\pi$.}
\label{fig2}
\end{figure*}
\begin{figure*}[ht]
\centering
\includegraphics[width=7cm]{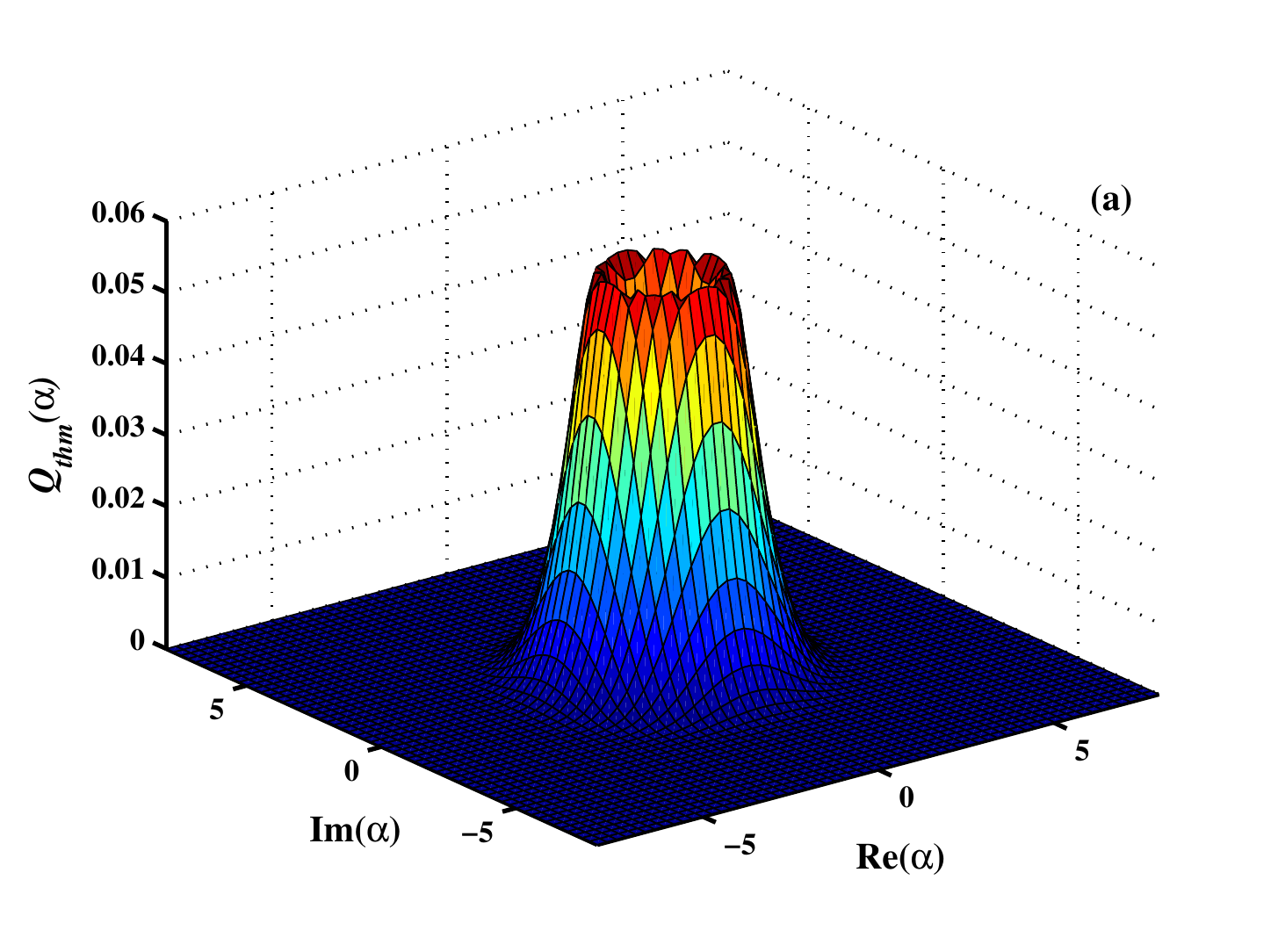}\hspace{2cm}
\includegraphics[width=7cm]{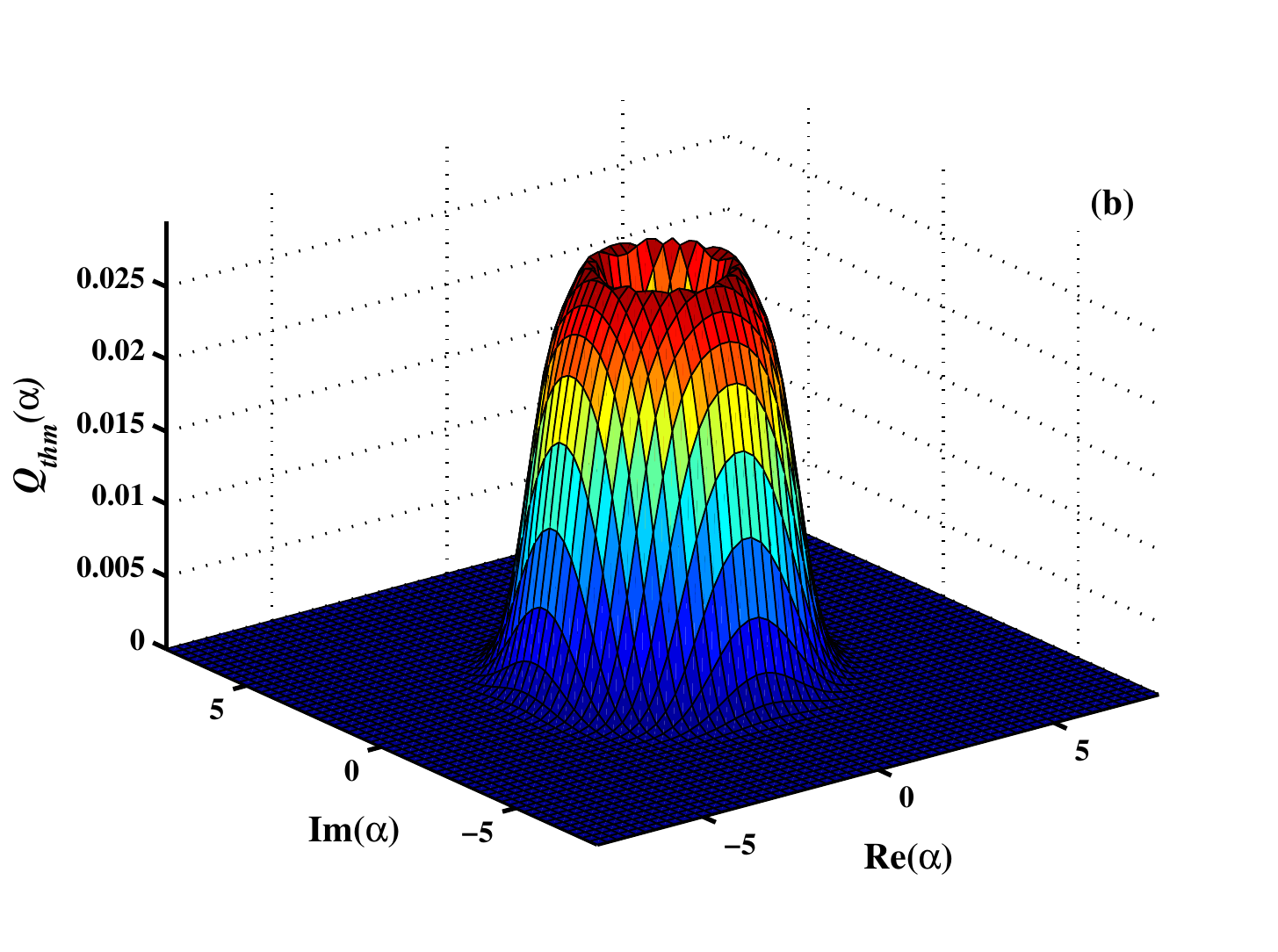}

\caption{(Color online) The $Q$-distribution for the initial thermal field is plotted against the parameters Re$(\alpha)$ and
Im$(\alpha)$ along $x$ and $y$ axes respectively with $gt_1=gt_2=2\pi/3$, (a) $\bar{n}=2$, (b) $\bar{n}=12$.}
\label{fig3}
\end{figure*}

In Fig.~\ref{fig2} and Fig.~\ref{fig3}, we have sketched the mesh plots of the $Q$ functions in the complex $\alpha$-plane if the atom
starts from the superposition state of $|e\rangle$ and $|i\rangle$ and the field is either in coherent or in thermal state respectively.

The mean photon number of the initial coherent state is taken as $|\alpha_0|^2=4$. Fig.~\ref{fig2}(a) shows that the $Q$ function initially represents
a simple Gaussian distribution centered around $(0, 0)$ whereas Fig.~\ref{fig2}(b) presents that in course of time the one-peaked $Q$ distribution
is divided into two peaks of similar amplitude but opposite phase. As time goes one of the peak has reduced its height with respect to the other
and then this truncated peak is fragmented into two small parts [see Figs.~\ref{fig2}(c) and (d)]. Thus Fig.~\ref{fig2} describes the deformation of the $Q$ function
for different values of $gt_1$ and $gt_2$.

Numerical result for Eq.~(\ref{eq11}) is presented in Fig.~\ref{fig3}, where we have plotted the $Q$ function of the initial thermal field as a function of $\alpha$. By taking a fixed interaction time $2\pi/3$, a hollowed-peak Gaussian structure is obtained. It is interesting to note that the increase of mean photon number of the thermal field does not change the shape of the $Q$ function much more. It slightly decreases the Gaussianity of the state but broadens the peak a bit.

\subsection{Wigner Distribution}

In this section, we analyze how the classical behavior of the coherent or the thermal field inside the cavity is affected by the passage of two identical atoms by considering the phase-space measure.
For any state having density matrix $\rho$ in the Fock state basis $\sum_{n, m}C_{n, m} |n\rangle\langle m|$, the Wigner function is defined by \cite{pathak05}
\begin{eqnarray}\nonumber
W(\alpha, \alpha^*) & = & \frac{2}{\pi^2} e^{2|\alpha|^2} \int{\langle-\gamma|\rho|\gamma\rangle e^{-2(\gamma\alpha^*-\gamma^*\alpha)}d^2\gamma},\\\nonumber
& = & \frac{2}{\pi} e^{2|\alpha|^2} \sum_{n, m}C_{n, m}\frac{(-1)^{n+m}}{2^{n+m}}\frac{\partial^{n+m}}
{\partial\alpha^n\partial\alpha^{*m}} e^{-4|\alpha|^2}.\\
\label{eq12}
\end{eqnarray}

\textit{Coherent state}: If the radiation field is initially in a coherent state $|\alpha_0\rangle$, then integral (\ref{eq12}) together with Eq.~(\ref{eq8}) yields the Wigner function as
\begin{eqnarray}\nonumber
W_{coh}(\alpha) & = & \frac{2}{\pi}e^{-(|\alpha_0|^2+2|\alpha|^2)}\left[\sum_{n, m}\frac{2^{n+m-2}}{(n-1)!(m-1)!}(\alpha_0\alpha^*)^{n-1}
(\alpha_0^*\alpha)^{m-1}\right.\\\nonumber
& & \left.\sin(\sqrt{2n}gt_1)\cos(\sqrt{2n}gt_2)\sin(\sqrt{2m}gt_1)\cos(\sqrt{2m}gt_2)\right.\\\nonumber
& & \left.+\sum_{n, m}\frac{2^{n+m}}{n!m!}(\alpha_0\alpha^*)^{n}(\alpha_0^*\alpha)^{m}\sin(\sqrt{2n}gt_1)\sin(\sqrt{2n}gt_2)\right.\\
& & \left.\sin(\sqrt{2m}gt_1)\sin(\sqrt{2m}gt_2)\right].
\label{eq13}
\end{eqnarray}
\begin{figure*}[ht]
\centering
\includegraphics[width=7cm]{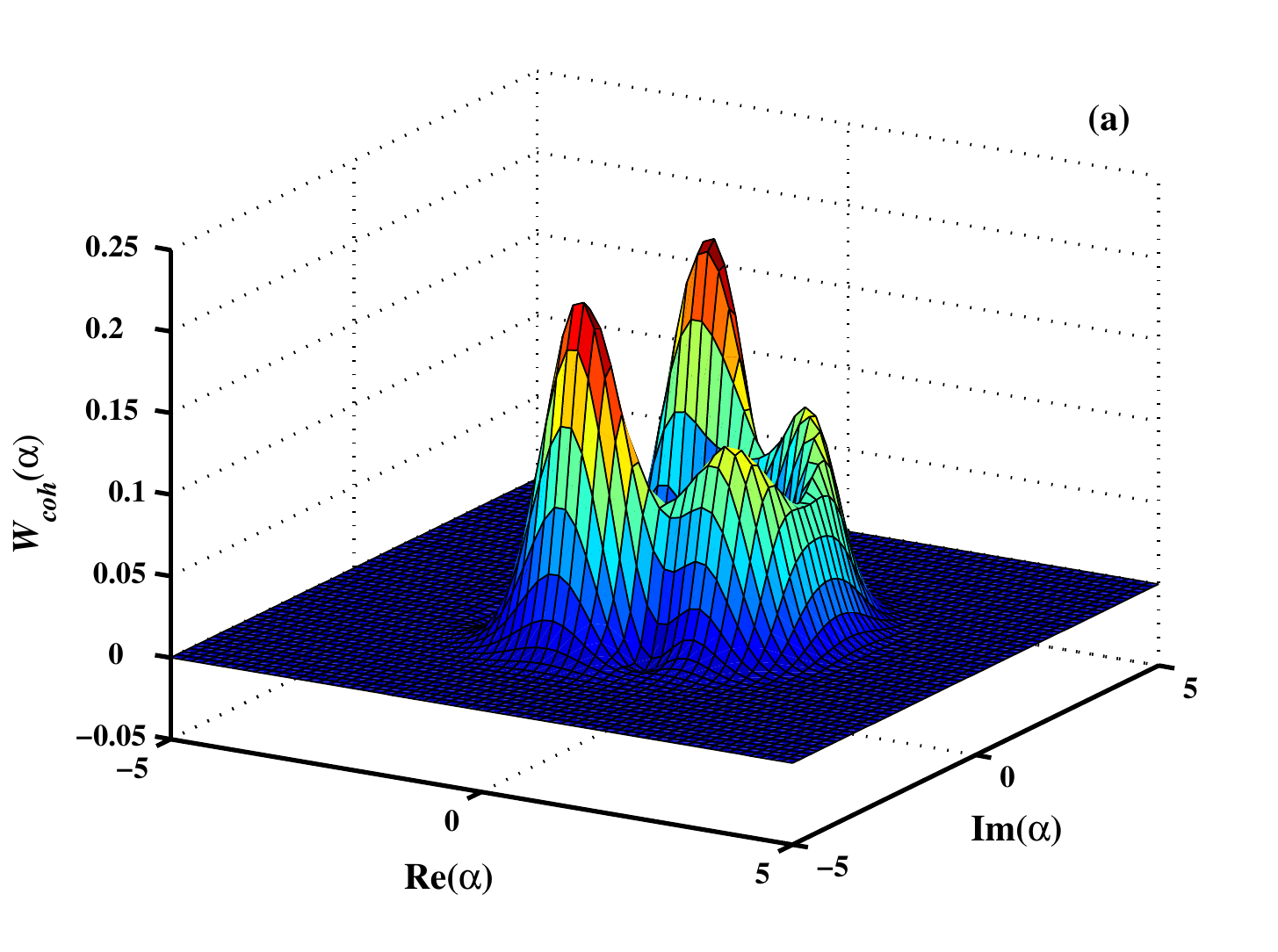}\hspace{2cm}
\includegraphics[width=7cm]{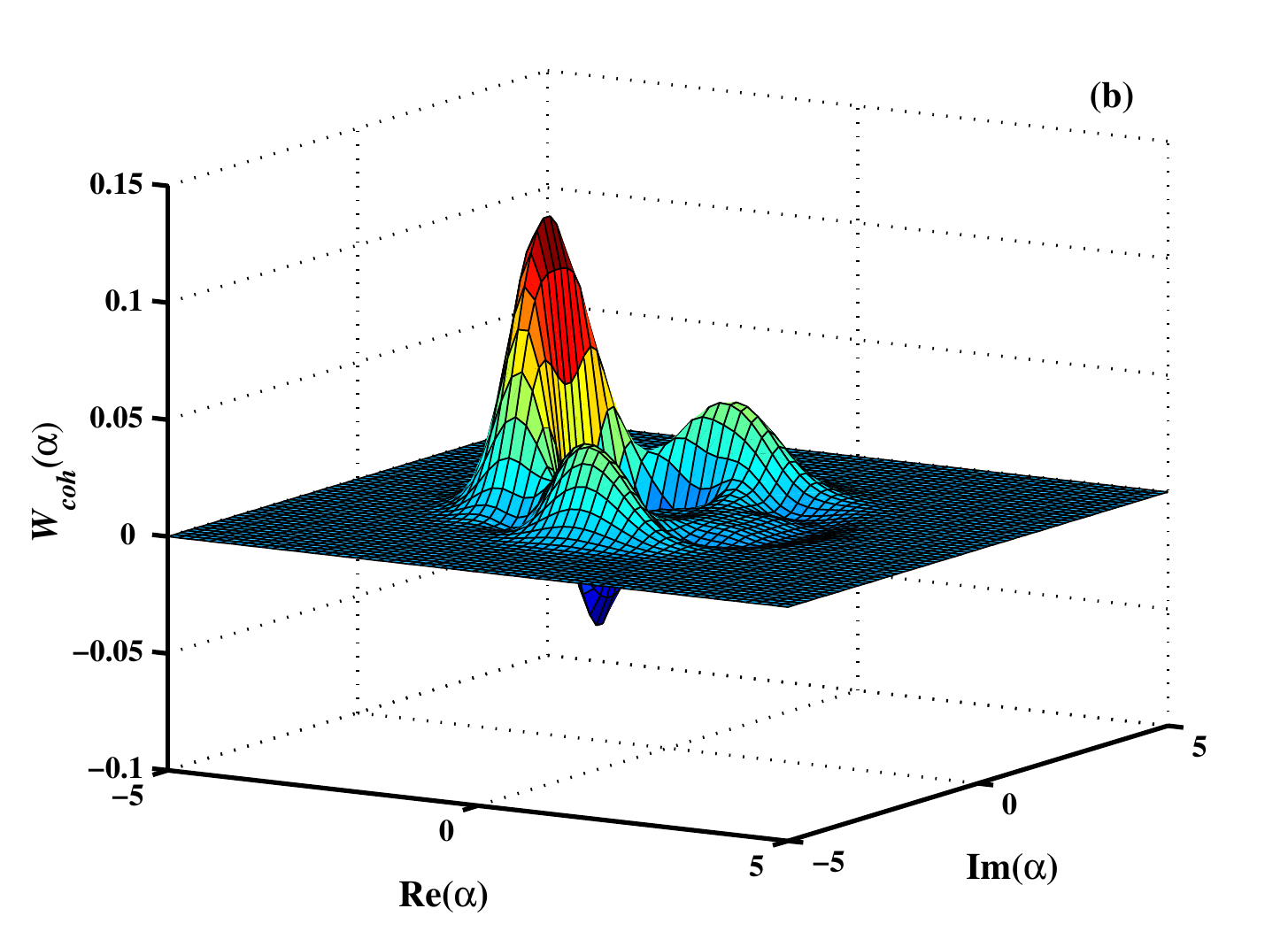}\hspace{2cm}
\includegraphics[width=7cm]{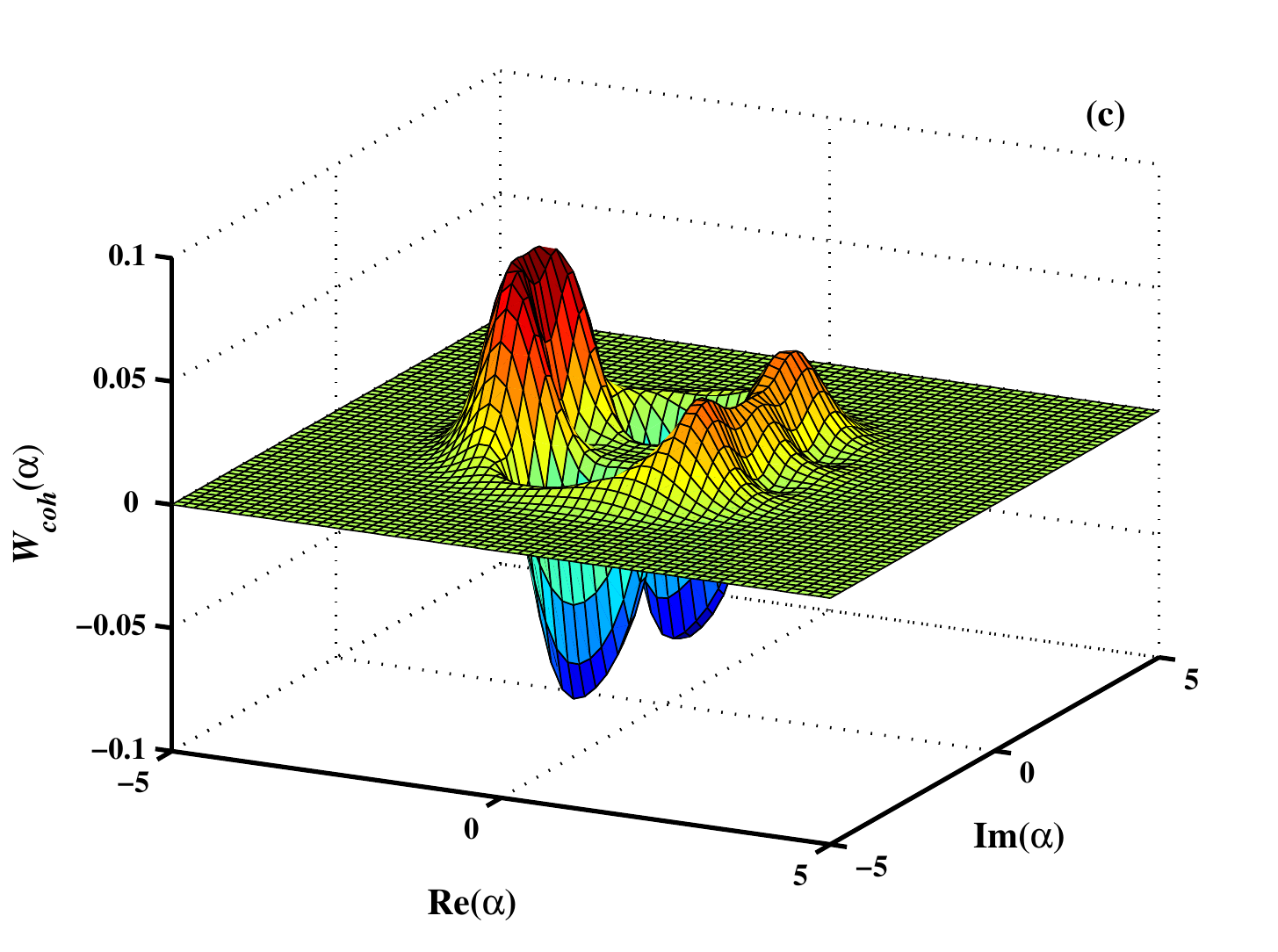}\hspace{2cm}
\includegraphics[width=7cm]{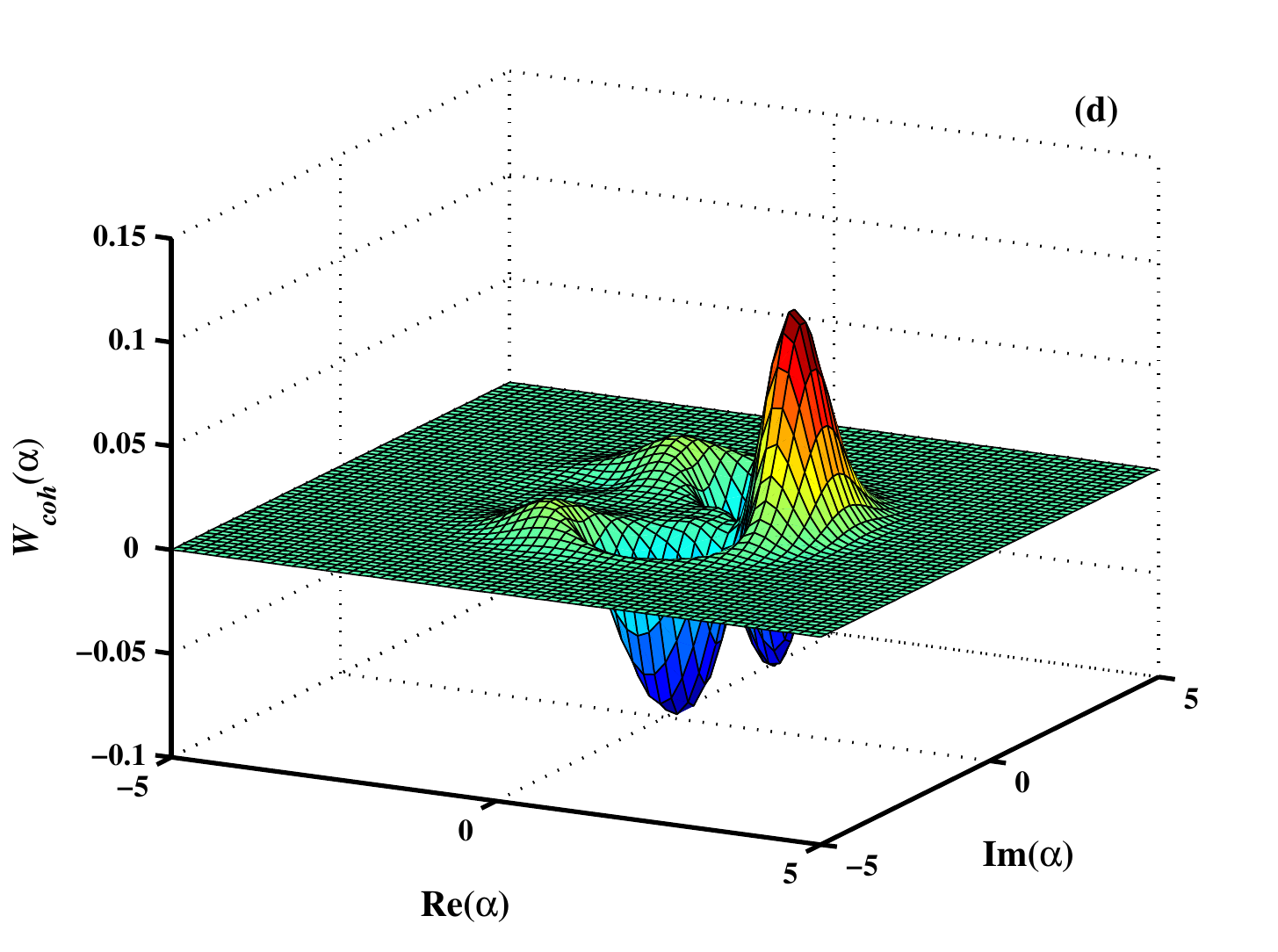}

\caption{(Color online) Three dimensional plot of $W(\alpha)$ when the field is initially in coherent state, using parameters $\alpha_0=2$
and (a) $gt_1=\pi/2$, $gt_2=3\pi/2$, (b) $gt_1=\pi/2$, $gt_2=5\pi/2$, (c) $gt_1=3\pi/2$, $gt_2=7\pi/2$ and
(d) $gt_1=5\pi/2$, $gt_2=7\pi/2$.}
\label{fig4}
\end{figure*}

Fig.~\ref{fig4} elaborates the Wigner function (\ref{eq13}) for a fixed $\alpha_0=2$ and for different values of $gt_1$ and $gt_2$.
It is known that the partial negativity of the Wigner function is a sufficient condition for tracing out the nonclassicality of quantum states.
In Fig.~\ref{fig4}(a), the negative part is slightly noticeable for $gt_1=\pi/2$ and $gt_2=3\pi/2$. The Wigner function includes more pronounced negative dips as time parameter changes from Fig.~\ref{fig4}(b) to Fig.~\ref{fig4}(d) and the positive multi-peak structure of the Wigner function disappears gradually. That means the classical nature of the initial coherent field is washed out by the crossing of two atoms in sequence.

\textit{Thermal state}: A thermal state $\rho_{thm}$ with average photon
number $\bar{n}$, acting as an input state, results the Wigner distribution
\begin{eqnarray}\nonumber
W_{thm}(\alpha) & = & \frac{2}{\pi}\frac{1}{(\bar{n}+1)}e^{-2|\alpha|^2}\left[\sum_n\left(\frac{4\bar{n}|\alpha|^2}
{\bar{n}+1}\right)^{n-1} \sin^2(\sqrt{2n}gt_1)\cos^2(\sqrt{2n}gt_2)\right.\\
& & +\left.\sum_n\left(\frac{4\bar{n}|\alpha|^2}{\bar{n}+1}\right)^n\sin^2(\sqrt{2n}gt_1)\sin^2(\sqrt{2n}gt_2)\right].
\label{eq14}
\end{eqnarray}
Eq.~(\ref{eq14}) is just the analytical expression of the Wigner function with mean thermal photon number $\bar{n}$. Unlike the coherent state input,
this function has no negative domain.

\section{Statistical properties}
\label{sec4}

Next we investigate two observable nonclassical effects, sub-Poissonian photon statistics and quadrature squeezing.
First to determine the photon statistics of a single-mode radiation field, we consider the Mandel's $Q$
parameter defined by \cite{mandel79}
\begin{eqnarray}
Q^M = \frac{\langle {a^\dag}^2 a^2\rangle-{\langle a^\dag a\rangle}^2}{\langle a^\dag a\rangle}.
\end{eqnarray}
For $-1\leq Q^M<0~(Q^M>0)$, the statistics is sub-Poissonian (super-Poissonian); $Q^M=0$ stands for Poissonian photon statistics.
To examine the statistical condition of the resulted cavity field, we obtain
\begin{eqnarray}\nonumber
\langle a^\dag a\rangle & = & 2\sum_n (n-1)|F_{n-1}|^2\sin^2(\sqrt{2n}gt_1)\cos^2(\sqrt{2n}gt_2)\\\nonumber
& & +\sum_n n|F_n|^2\sin^2(\sqrt{2n}gt_1)\sin^2(\sqrt{2n}gt_2),
\end{eqnarray}
and
\begin{eqnarray}\nonumber
\langle {a^\dag}^2 a^2\rangle & = & 2\sum_n (n-2)(n-1)|F_{n-1}|^2\sin^2(\sqrt{2n}gt_1)\cos^2(\sqrt{2n}gt_2)\\\nonumber
& & +\sum_n (n-1)n|F_n|^2\sin^2(\sqrt{2n}gt_1)\sin^2(\sqrt{2n}gt_2),
\end{eqnarray}
where ${|F_n|}^2$ stands for the initial photon distribution.

\begin{figure}[h]
\centering
\includegraphics[width=6cm]{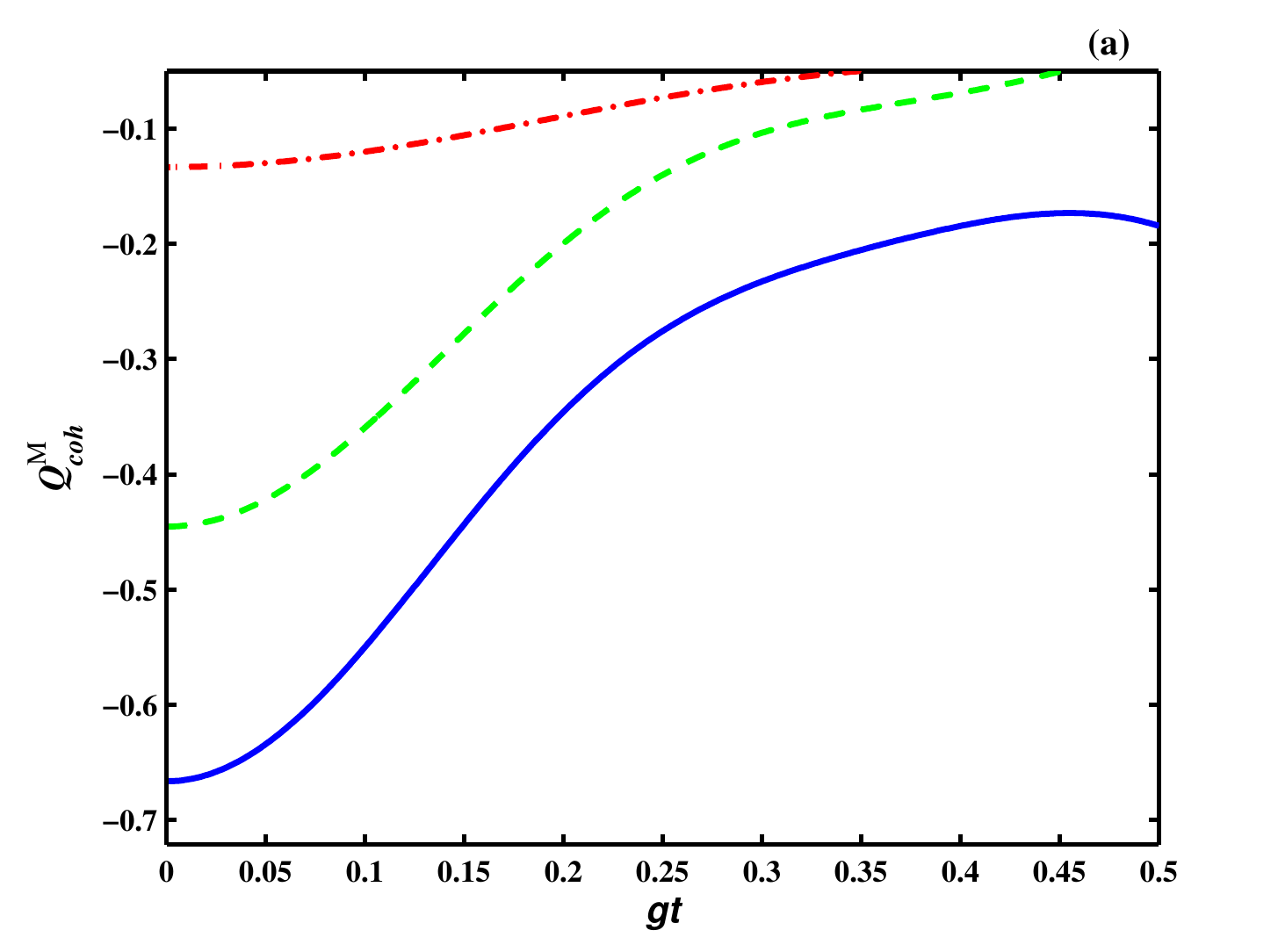}
\includegraphics[width=6cm]{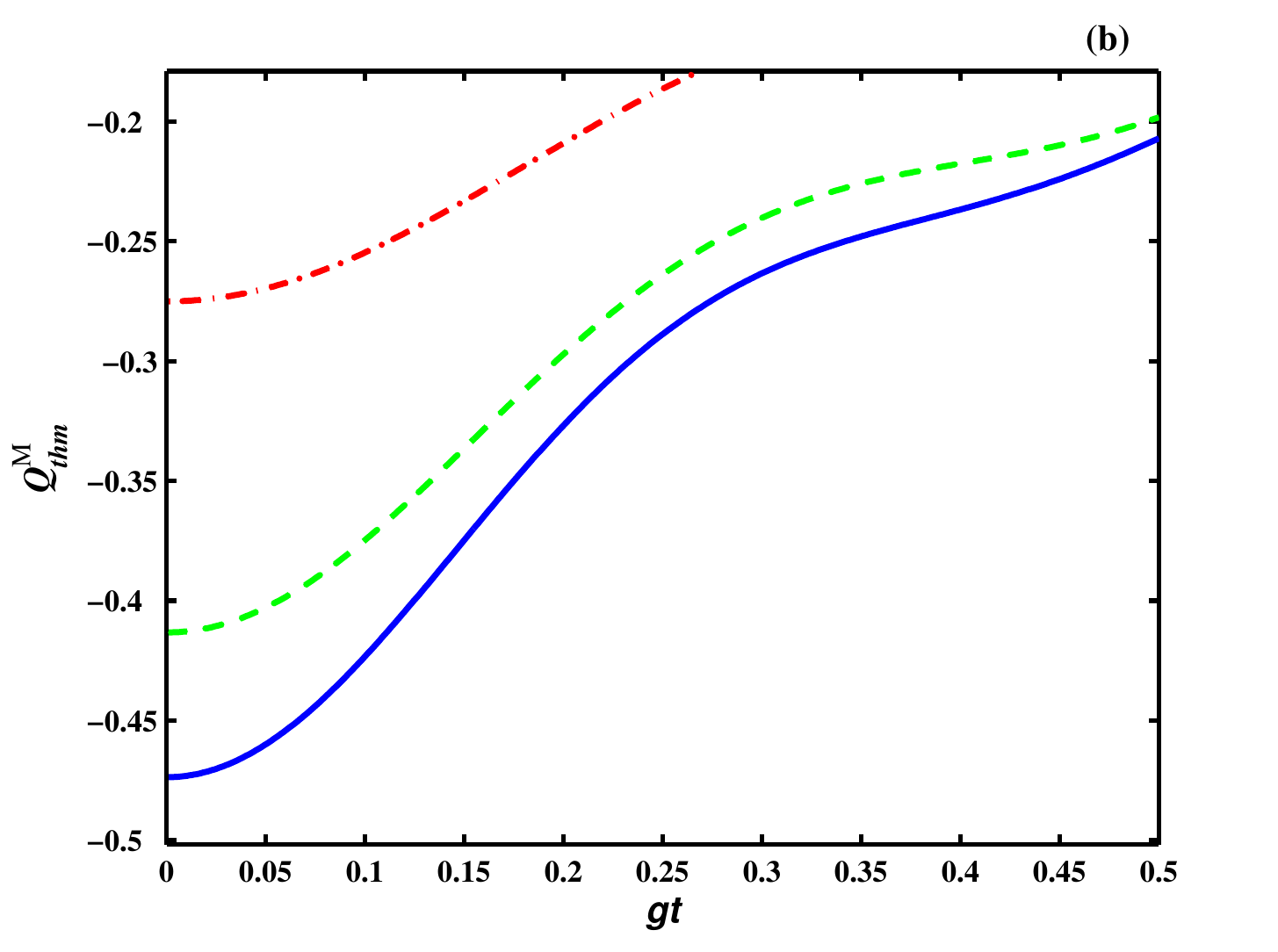}

\caption{(Color online) Mandel's $Q^M$ as a function of $gt_1=gt_2=gt$ and with (a) $\alpha_0=1$, 2 and 3 (from upper to lower curves) for a coherent state input and (b) $\bar{n}=1$, 2, 3 (from upper to lower curves) for a thermal state input.}
\label{fig5}
\end{figure}

In order to see the variation of the $Q^M$ parameter with $\alpha_0$ (coherent field) or $\bar{n}$ (thermal field), we plot the $Q^M$ function against the scaled time $gt$ in Fig.~\ref{fig5}. $Q^M$ always exhibits sub-Poissonian character for both the input states and increases its negativity as $\alpha_0\,(\bar{n})$ increases. This implies that the nonclassicality is enhanced by increasing the initial photon number. We should emphasize that though the thermal field has no negative Wigner function but it displays the sub-Poissonian property.


Secondly, to analyze the squeezing properties of the radiation field we introduce two hermitian quadrature operators
\begin{eqnarray}
X=a+a^\dag,~~~~~~Y=-i(a-a^\dag).
\end{eqnarray}
These two quadrature operators satisfy the commutation relation $[X, Y]=2i$ and, as a result, the uncertainty principle $(\Delta X)^2(\Delta Y)^2\geq 1$.
A state is said to be squeezed if either $(\Delta X)^2$ or $(\Delta Y)^2$ is less than 1. To review the principle of quadrature squeezing \cite{luks88}, we define
an appropriate quadrature operator \cite{wang03}
\begin{eqnarray}
X_\theta = X\cos\theta+Y\sin\theta = ae^{-i\theta}+a^\dag e^{i\theta}.
\end{eqnarray}
The squeezing of $X_\theta$ is characterized by the condition
$\langle:(\Delta X_\theta)^2:\rangle<0$ where the double dots denote the normal ordering of
operators. After expanding the terms of $\langle:(\Delta X_\theta)^2:\rangle$ and minimizing its value over the whole angle $\theta$, one can get \cite{lee10}
\begin{eqnarray}\nonumber
S_{opt} & = & \langle:(\Delta X_\theta)^2:\rangle_{min}\\
& = & -2|\langle a^{\dag 2}\rangle-\langle a^\dag\rangle^2|+2\langle a^\dag a\rangle-2|\langle a^\dag \rangle|^2.
\label{eq18}
\end{eqnarray}
For the atomic system under consideration, $\langle a^\dag a\rangle$ has been derived earlier and the other parameters of Eq.~(\ref{eq18})
are given as
\begin{eqnarray}\nonumber
\langle a^\dag\rangle & = & 2\sum_n \sqrt{n}~F_{n-1}\overline{F}_{n+1}\sin(\sqrt{2n-2}gt_1)\sin(\sqrt{2n+2}gt_2)\cos(\sqrt{2n-2}gt_1)\\\nonumber
& & \cos(\sqrt{2n+2}gt_2)+\sum_n \sqrt{n+1}~F_{n}\overline{F}_{n+2}\sin(\sqrt{2n-2}gt_1)\sin(\sqrt{2n-2}gt_2)\\\nonumber
& &\sin(\sqrt{2n+4}gt_1)\sin(\sqrt{2n+4}gt_2)
\end{eqnarray}
and
\begin{eqnarray}\nonumber
\langle a^{\dag 2}\rangle & = & 2\sum_n \sqrt{n}\sqrt{n+1}~F_{n-1}\overline{F}_{n+2}\sin(\sqrt{2n-2}gt_1)\sin(\sqrt{2n+4}gt_2)\cos(\sqrt{2n-2}gt_1)\\\nonumber
& & \cos(\sqrt{2n+4}gt_2)+\sum_n \sqrt{n+1}\sqrt{n+2}F_{n}\overline{F}_{n+3}\sin(\sqrt{2n-2}gt_1)\sin(\sqrt{2n-2}gt_2)\\\nonumber
& &\sin(\sqrt{2n+6}gt_1)\sin(\sqrt{2n+6}gt_2),
\end{eqnarray}
Substituting the above expectation values in Eq.~(\ref{eq18}) we obtain lengthy expressions of $S_{opt}$ for the initial coherent (thermal) state when $F_n=e^{-|\alpha_0|^2/2}\frac{{\alpha_0}^n}{\sqrt{n!}}$ $\left(\sum \frac{{\bar{n}}^n}{(\bar{n}+1)^{n+1}}\right)$.

\begin{figure}[h]
\centering
\includegraphics[width=6cm]{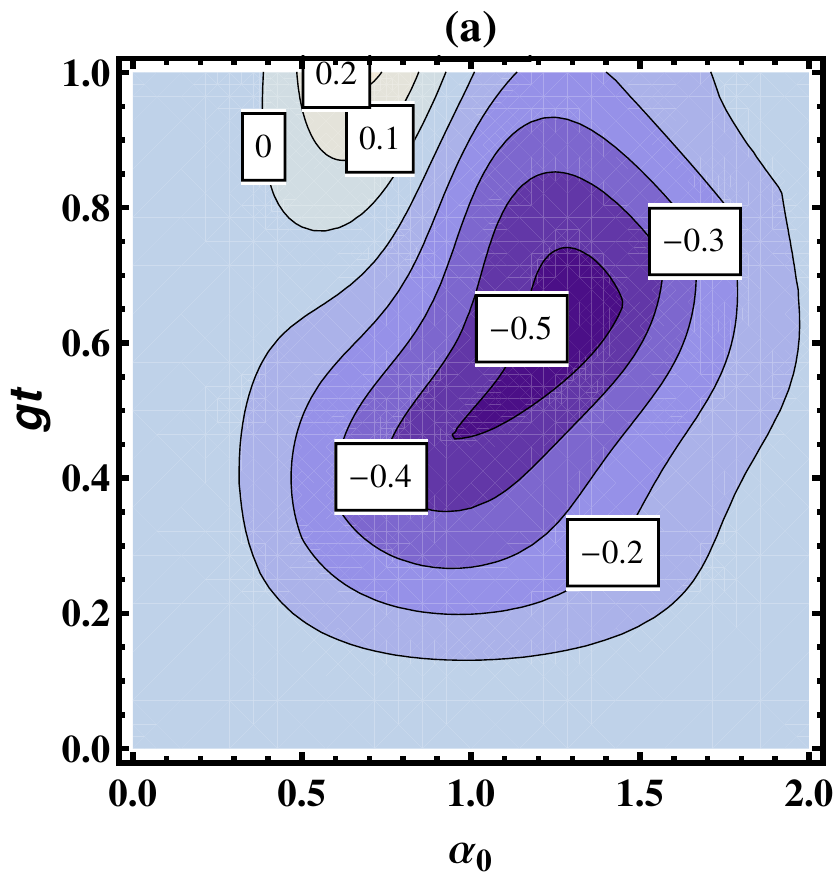}
\includegraphics[width=6cm]{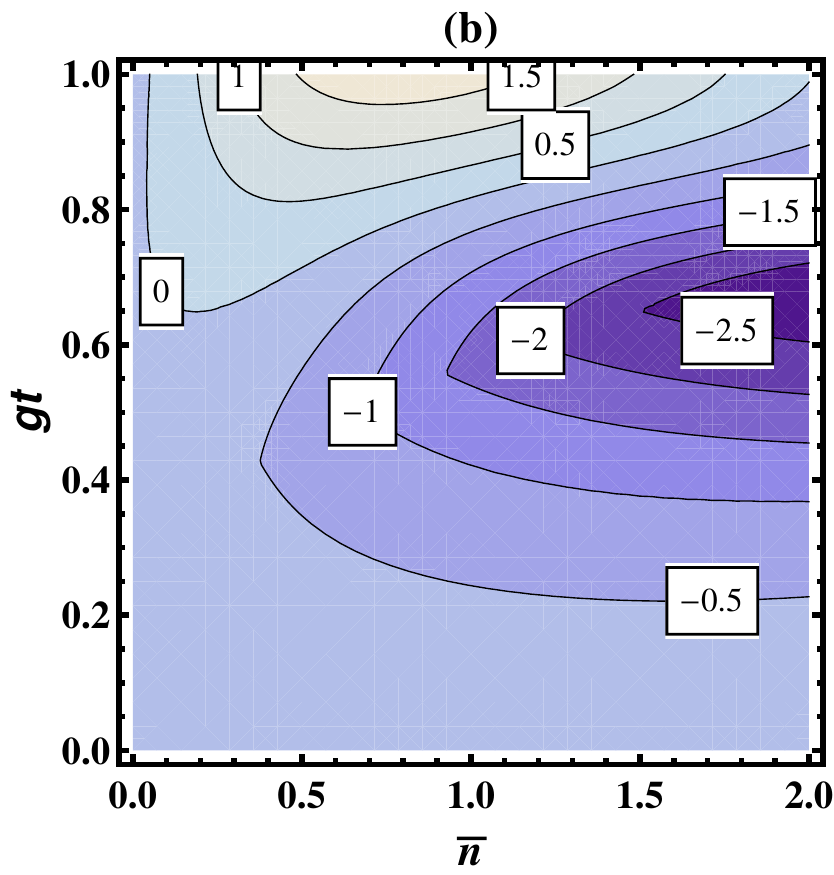}

\caption{(Color online) Contour plot for $S_{opt}$ as a function of $gt_1=gt_2=gt$ and (a) $\alpha_0$ (coherent state) and (b) $\bar{n}$ (thermal state).}
\label{fig6}
\end{figure}

Fig.~\ref{fig6}(a) presents the contour plot of $S_{opt}$ as a function of $\alpha_0$ and $gt$. One can see clearly that
the traveling of two atoms through the cavity inject the squeezing property into the coherent state character. In addition, the field transferred from the initial thermal field also depicts squeezing effect [see Fig.~\ref{fig6}(b)].

\section{Conclusion}
\label{sec5}

In this article, we have proposed to fly two three-level atoms one after one through the cavity field. Incorporating zero detuning assumption in the atom-field coupling, an analytical expression for the quasi $Q$-distribution is derived when the field is initially prepared either in a coherent or in a thermal state. It has been shown that the movement of atoms slightly changes the shape of the quasiprobability $Q$ function for both the input states. Furthermore, the nonclassicality of the resulted field is discussed in terms of the negativity of the Wigner function, Mandel's $Q$ parameter and the quadrature squeezing. The Wigner function of the coherent state input always exhibits partially negative region which is a clear evidence of its nonclassical behavior. Figs.~\ref{fig4}(a)-(d) demonstrate that the negativity of $W$ function gradually increases with time. As a consequence of the nonclassicality, the coherent field depicts sub-Poissonian photon-number distribution and quadrature squeezing.

In addition, Wigner function does not always indicate a negative value for the nonclassical state, e.g. the squeezed state is usually considered as a typical nonclassical state since its quadrature noise is less than that of the vacuum, but its Wigner function is regular and positive \cite{filip11}. While for the thermal description of the initial field, we have obtained an almost similar case. The resulted field owns a positive Wigner function but exhibits sub-Poissonian photon statistics and squeezing. In particular, the output state with the input coherent (thermal) state achieves better sub-$Q$ property for better coherent amplitude $\alpha_0$ (average photon number $\bar{n})$.


\begin{center}
\textbf{Acknowledgement}
\end{center}
AC thanks National Board of Higher Mathematics, Department of Atomic Energy, India for the financial support.

\end{document}